\documentclass[twocolumn,showpacs,prd,nofootinbib]{revtex4}
\usepackage{amssymb}
\usepackage{enumerate}
\usepackage{dcolumn}

\usepackage{graphicx}
\usepackage{dcolumn}
\usepackage{bm}

\begin{document}


\def\H{{\cal H}}
\def\ttheta{\tilde{\theta}}

\def\beq{\begin{equation}}
\def\eeq{\end{equation}}
\def\bea{\begin{eqnarray}}
\def\eea{\end{eqnarray}}
\def\ben{\begin{enumerate}}
\def\een{\end{enumerate}}
\def\la{\langle}
\def\ra{\rangle}
\def\a{\alpha}
\def\b{\beta}
\def\g{\gamma}\def\G{\Gamma}
\def\d{\delta}
\def\e{\epsilon}
\def\phi{\varphi}
\def\k{\kappa}
\def\l{\lambda}
\def\m{\mu}
\def\n{\nu}
\def\o{\omega}
\def\p{\pi}
\def\r{\rho}
\def\s{\sigma}
\def\t{\tau}
\def\L{{\cal L}}
\def\S{\Sigma }
\def\gsim{\; \raisebox{-.8ex}{$\stackrel{\textstyle >}{\sim}$}\;}
\def\lsim{\; \raisebox{-.8ex}{$\stackrel{\textstyle <}{\sim}$}\;}
\def\gtrsim{\gsim}
\def\lessim{\lsim}
\def\loc{{\rm local}}
\def\vm{v_{\rm max}}
\def\bh{\bar{h}}
\def\del{\partial}
\def\nab{\nabla}
\def\half{{\textstyle{\frac{1}{2}}}}
\def\fourth{{\textstyle{\frac{1}{4}}}}

\title{Numerical simulations of gravitational collapse in
Einstein-aether theory}

\author{David Garfinkle}
\email{garfinkl@oakland.edu}
\affiliation{Department of Physics, Oakland University, Rochester, MI 48309}

\author{Christopher Eling}
\email{cteling@physics.umd.edu}
\affiliation{Department of Physics, University of Maryland, College Park, MD
20742}

\author{Ted Jacobson}
\email{jacobson@umd.edu} \affiliation{Department of Physics,
University of Maryland, College Park, MD 20742}

\begin{abstract}

We study gravitational collapse of a spherically symmetric scalar
field in Einstein-aether theory (general relativity coupled to a
dynamical unit timelike vector field). The initial value formulation
is developed, and numerical simulations are performed. The collapse
produces regular, stationary black holes, as long as the aether
coupling constants are not too large. For larger couplings a finite
area singularity occurs. These results are shown to be consistent
with the stationary solutions found previously.

\end{abstract}
\maketitle

\section{Introduction}
Over the past several years there has been a renewal of interest in
the physics of Lorentz symmetry violation. These investigations are
largely motivated by the enduring problem of quantum gravity and
also cosmological mysteries such as the dark matter and dark energy
problems. Lorentz violation (LV) effects in Standard Model
extensions and modified particle dispersion relations are currently
tightly constrained by high precision tests \cite{Mattingly:2005re},
however constraints on gravitational LV effects are not as severe.
Various approaches to incorporating LV into gravity have been
investigated, see for example Refs.
~\cite{LVmetr,Gasperini:1986ym,Kostelecky:1989jw,Clayton:1999zs,
Jacobson:2000xp, Arkani-Hamed:2003uy,Bekenstein:2004ne,
Gripaios:2004ms,Bluhm:2004ep, Heinicke:2005bp,Rubakov:2006pn,
Cheng:2006us,Zlosnik:2006zu}.

In this paper we will consider ``Einstein-aether" theory (or
ae-theory, for short) \cite{Jacobson:2000xp,Eling:2004dk}, which
consists of a dynamical unit timelike vector field $u^a$ coupled to
Einstein gravity. The $u^a$ ``aether" field can be thought of as the
4-velocity of a preferred frame. The LV here is a form of
spontaneous symmetry breaking, since  the unit timelike constraint
requires the vector field to have a non-zero value everywhere in any
solution, even flat spacetime.

The Lagrangian of ae-theory depends on four dimensionless coupling
constants $c_{1,2,3,4}$. By studying the phenomenology of ae-theory
it can be compared with observations in order to constrain the $c_i$
values and possibly rule the theory out entirely as a description of
nature. Observational constraints on ae-theory have been determined
from parameterized post-Newtonian analysis~\cite{Eling:2003rd,
Graesser:2005bg,Foster:2005dk}, stability and energy
positivity~\cite{Jacobson:2004ts,Lim:2004js, Elliott:2005va,
Eling:2005zq}, primordial nucleosynthesis~\cite{Carroll:2004ai}, and
vacuum Cerenkov radiation~\cite{Elliott:2005va}. The combined
constraints from all of these are reviewed in
Ref.~\cite{Foster:2005dk}. They are all consistent with a large
(order unity) 2d region in the 4d parameter space. Agreement with
weak field radiation damping (such as in binary pulsar
systems)~\cite{Foster:2006az, Fosternext} is expected to restrict
the parameters to a narrow band of width $\sim 10^{-3}$ in that 2d
region, and combined constraints from many such systems may further
narrow the allowed region to a small neighborhood of the origin,
with size of order $10^{-1}$-$10^{-2}$.

It is likely that theories with LV can generally be more tightly
constrained by examining strong field solutions. In addition, the
properties of such solutions may also be of interest from a purely
mathematical physics perspective since the additional fields that
couple to GR can produce interesting new sorts of solutions and
phenomena. Black hole solutions have been considered in the ghost
condensate theory \cite{Mukohyama:2005rw}, and spherically symmetric
solutions in the relativistic MOND theory were studied in
\cite{Giannios:2005es}. Work on this subject in ae-theory was begun
long ago in special cases
~\cite{Gasperini:1986ym,Kostelecky:1989jw,Jacobson:2000xp}, and a
study of the solutions for general values of the coupling parameters
was completed recently in \cite{Eling:2006df} and
\cite{Eling:2006ec}.

In these papers the time independent spherically symmetric solutions
in ae-theory were characterized by examining the field equations as
a set of coupled ordinary differential equations. It was
demonstrated that there is generically a three parameter family of
solutions to these equations. Imposing the boundary condition of
asymptotic flatness at spatial infinity reduces the solution space
to a two parameter family. This differs from GR, in which there is a
unique spherically symmetric solution, the Schwarzschild metric,
which is automatically asymptotically flat. The one parameter subset
of solutions where $u^a$ is aligned with the timelike Killing vector
was found analytically and discussed in detail in
\cite{Eling:2006df}. It was also shown that this solution describes
the exterior of a static fluid star.

Regular black holes in ae-theory must be described by a different
solution, since the aether is a timelike vector so it cannot
coincide with the Killing vector which is null on the horizon and
spacelike inside. That is, the aether must be flowing into the black
hole. Moreover, the theory possesses spin-2, spin-1, and spin-0
``massless" wave modes that travel at speeds that differ from one
another and from the metric speed of light~\cite{Jacobson:2004ts}.
To qualify as a black hole, a solution must therefore contain a
region where all of these wave modes are trapped. It was found
in~\cite{Eling:2006ec} that in spherical symmetry, regularity at the
metric, spin-2, and spin-1 horizons is automatic, but regularity at
the spin-0 horizon is a supplementary condition. This condition
reduces the two-parameter family of asymptotically flat solutions to
a one-parameter family characterized by the total mass or horizon
area. Particular black hole solutions with regular spin-0 horizons
were found by numerical integration for special cases of the
coupling parameters. Like the Schwarzschild solution, these
solutions have spacelike singularities at the origin, and in the
exterior the metric functions are very close to the familiar
Schwarzschild solution, even for fairly large coupling constants
$c_i$. (The differences become more pronounced in the black hole
interior.) For certain large values of the coupling constants no
black hole solution with a regular spin-0 horizon was found.

{}From an astrophysical point of view,  an important question is which
solutions describe the final state of a gravitational collapse of
matter fields. In \cite{Eling:2006ec} it was conjectured that the
solutions with regular spin-0 horizons---when they exist---are the
ones that arise from collapse of initially nonsingular data. The
main objective of this paper is to test that conjecture by numerical
evolution of the time dependent spherically symmetric field
equations given appropriate collapsing initial data. A massless
scalar field coupled only to the metric will serve as the matter
field.

In Section \ref{aetheory} we review the nature and
status of Einstein-aether theory, and discuss the
nature of black holes in this theory. Section \ref{IVform}
together with the Appendix develops the initial value
formulation under the assumption that the aether vector is
hypersurface orthogonal, as is always the case in
spherical symmetry.  The results of numerical collapse
simulations are presented in Section \ref{Numresults}.
We find that the collapsing scalar field does indeed reach a
final stationary black hole state with a regular spin-0 horizon, for
the range of coupling constants found  in \cite{Eling:2006ec} to
allow such states. Outside this range of $c_i$ we find solutions
that develop a finite area singularity at the would-be spin-0
horizon. The numerical evolution procedure also allows for wider
explorations of space of $c_i$ than was done
in~\cite{Eling:2006ec}. The results appear to support the conjecture
that formation of black holes with regular spin-0 horizons is
generic in ae-theory for coupling parameters that are not too large.

\section{Einstein-aether theory}
\label{aetheory}

The action for Einstein-aether theory is the most general generally
covariant functional of the spacetime metric $g_{ab}$ and aether
field $u^a$ involving no more than two derivatives (not including
total derivatives),
\beq S = \int \sqrt{-g}~ (L_{\rm ae}+L_{\rm matter})
~d^{4}x \label{action} \eeq
where
\beq L_{\rm ae} =  \frac{1}{16\pi G} [R-K^{ab}{}_{mn} \nabla_a u^m
\nabla_b u^n +
\lambda(g_{ab}u^a u^b + 1)] \eeq
and $L_{\rm matter}$ denotes the matter lagrangian. Here $R$ is
the Ricci scalar, ${K^{ab}}_{mn}$ is defined as
\beq {{K^{ab}}_{mn}} = c_1 g^{ab}g_{mn}+c_2\delta^{a}_{m} \delta^{b}_{n}
+c_3\delta^{a}_{n}\delta^{b}_{m}-c_4u^a u^b g_{mn} \eeq
where the $c_i$ are dimensionless coupling constants, and $\lambda$
is a Lagrange multiplier enforcing the unit timelike constraint on
the aether. The convention used in this paper for metric signature
is $({-}{+}{+}{+})$ and the units are chosen so that the speed of
light defined by the metric $g_{ab}$ is unity. In the weak-field,
slow-motion limit ae-theory reduces to Newtonian
gravity~\cite{Carroll:2004ai}, with a value of  Newton's constant
$G_{\rm N}$ related to the parameter $G$ in the action
(\ref{action})  by $G_{\rm N}=G(1-(c_1+c_4)/2)^{-1}$.

In this paper we take the matter field to be a minimally coupled
massless scalar field $\chi$, with Lagrangian $ - {\nabla _a}\chi
{\nabla ^a}\chi$.  However, to simplify the form of the equations of
motion, we introduce the quantity $ \psi = \chi \sqrt{16 \pi G}$ so
that the matter Lagrangian takes the form
\beq L_{\rm matter} = \frac{-1}{16\pi G} \nabla_a \psi \nabla^a
\psi. \eeq

The field equations from varying (\ref{action}) with respect to
$g^{ab}$, $u^a$, $\psi$, and $\lambda$ are given respectively by
\bea
{G_{ab}} &=& {T_{ab}}
\label{EFE}
\\
{\nabla _a} {{J^a}_b} + \lambda {u_b} + {c_4} {a_a} {\nabla _b} {u^a} &=& 0
\label{evolveu}
\\
{\nabla ^a}{\nabla _a} \psi &=& 0
\\
{u^a}{u_a} &=& -1.\label{unit}
 \eea Here $G_{ab}$ is the Einstein
tensor of the metric $g_{ab}$; and the quantities ${J^a}_b,\; {a_a}$
and the stress-energy $T_{ab}$ are given by \bea {{J^a}_m} &=&
{{K^{ab}}_{mn}}{\nabla_b}{u^n}
\\
{a_a} &=& {u^b}{\nabla _b}{u_a}
\\
\nonumber
{T_{ab}} &=& - {\textstyle {1 \over 2}} {g_{ab}}\left ( {{J^c}_d}{\nabla_c}{u^d}
+ {\nabla _c}\psi {\nabla ^c} \psi \right )
\\
\nonumber
&+& {\nabla _a} \psi {\nabla _b} \psi + {c_4} {a_a}{a_b}
+ \lambda {u_a}{u_b}
\\
\nonumber
&+& {c_1} \left ( {\nabla_a}{u_c}{\nabla_b}{u^c} - {\nabla^c}{u_a}
{\nabla_c}{u_b}
\right )
\\
&+& {\nabla _c} \left [ {{J^c}_{(a}}{u_{b)}} + {u^c}{J_{(ab)}} -
{{J_{(a}}^c}{u_{b)}}\right ]. \label{aestress} \eea

In this paper we will perform numerical simulations of this system
in spherical symmetry.  The condition of spherical symmetry
implies that the twist ${\omega _a}={\epsilon _{abcd}}{u^b}{\nabla
^c}{u^d}$ of the vector field $u^a$ vanishes.  This makes the four
parameters of the Lagrangian redundant: the action depends only on
${c_2},\, {c_{13}}$ and $c_{14}$ where we use the abbreviation
$c_{13}$ to stand for ${c_1}+{c_3}$ (and correspondingly for
$c_{14}$). (In Ref.~\cite{Eling:2006ec} it is shown that when the
twist vanishes the $c_4$ term can absorbed by the shifts
$c_1\rightarrow c_1+c_4$ and $c_3\rightarrow c_3-c_4$. Thus the
action depends only on $c_2$, $c_{14}$ and $c_3-c_4$. But since
the latter is equal to $c_{13}-c_{14}$, the action is also
determined by $c_2$, $c_{13}$, and $c_{14}$.) Because the twist
vanishes, it follows that $u^a$ is hypersurface orthogonal.

For our study of scalar field collapse we considered a case where
the parameters $c_i$ satisfy the requirements that (i) the
Einstein-aether theory has precisely the same post-Newtonian
parameters as general relativity, so it agrees in the post-Newtonian
approximation with all experimental tests of general relativity,
(ii) the theory is linearly stable, (iii) the wave modes all carry
positive energy, and (iv) the wave modes travel at greater than or
equal to the matter speed of light, so there is no vacuum
\v{C}erenkov radiation. Together these impose the
conditions~\cite{Foster:2005dk} \bea {c_2} &=& {{c_3 ^2}\over {3
{c_1}}} - {{2{c_1}+{c_3}} \over 3}
\\
{c_4} &=& {{- {c_3 ^2}} \over {c_1}}
\\
0 &<& {c_{13}} < 1
\\
0 &<& ({c_1}-{c_3}) < {{c_{13}} \over {3(1-{c_{13}})}}. \eea
We chose
${c_1}=1/3$ and ${c_3}=1/6$ (essentially in the midrange of the
inequalities given above) and then determined
$c_2$ and $c_4$ by the
above equalities. This yields the parameter set
\beq
c_1=\frac{1}{3},\quad
c_2=-\frac{1}{4},\quad
c_3=\frac{1}{6},\quad
c_4=-\frac{1}{12}.
\label{ci}
\eeq

This choice of the constants $c_i$ is different from those used in
the studies of static black holes in \cite{Eling:2006ec}. However,
for the Einstein-aether theory without matter, the transformation
${u_a} \to {\sqrt \sigma} {u_a}$ (with $\sigma $ a constant and with
$h_{ab}$ remaining unchanged) takes Einstein-aether theory into
itself but with changes in the parameters
$c_i$~\cite{Foster:2005ec}. This transformation and the restriction
to spherical symmetry allows one to transform the results with our
parameter choice (\ref{ci}) to one with ${c_{13}}$ and $c_4$
vanishing and with ${c_1}=1/4$ and ${c_2}=1/2$. This new set of
parameters falls in the class that lends itself to a numerical
integration of the static, spherically symmetric Einstein-aether
equations using the methods of \cite{Eling:2006ec}, and thus allows
a direct comparison of a static solution with the end state of our
numerical simulations of gravitational collapse.

In this paper we examine the collapse process to determine the
nature of the black hole that is formed.  This issue is somewhat
more subtle in Einstein-aether theory than in general relativity.
Usually in a gravitational collapse simulation one looks for
apparent horizons: marginally outer trapped surfaces of the metric
$g_{ab}$.  Typically the spacetime outside the apparent horizon
settles down to a stationary (static in the spherically symmetric
case) black hole. However, Einstein-aether theory contains a spin-0
mode that travels at the speed $v_0$ where~\cite{Jacobson:2004ts}
\beq {v_0 ^2} = {{{c_{123}}(2-{c_{14}})} \over {{c_{14}}(1-{c_{13}})
(2+{c_{13}}+3{c_2})}}. \label{v0}\eeq
For the equalities that we have imposed
on the $c_i$ this expression reduces to
\beq {v_0 ^2} = {{c_{13}}\over {3 ({c_1}-{c_3})(1-{c_{13}})}}.
\label{v02} \eeq
Thus, the
boundary of the black hole should be a horizon that just barely
traps the spin-0 mode.  This turns out to be equivalent to a
marginally outer trapped surface of the metric ${{\tilde
g}_{ab}}={g_{ab}}+(1-{v_0 ^2}){u_a}{u_b}$. Thus in searching for
black holes we look for marginally outer
trapped
surfaces of the metric ${\tilde g}_{ab}$, which for brevity we will
call spin-0 horizons.
For the parameter choice (\ref{ci}),
the spin-0 speed from (\ref{v02}) is
$\sqrt{2}$, so the spin-0 horizon lies inside the metric horizon.

\section{Initial value formulation}
\label{IVform}

Numerical
evolution requires
a choice of the surfaces of constant time, and we will choose
those surfaces orthogonal to $u^a$.
This choice is always
possible in spherical symmetry since
$u^a$ is necessarily hypersurface orthogonal.
The spatial metric $h_{ab}$ and extrinsic
curvature $K_{ab}$ of these hypersurfaces are then given by
\bea
{h_{ab}} = {g_{ab}} + {u_a}{u_b}
\label{hdef}
\\
{K_{ab}} = - \half {{\cal L}_u} {h_{ab}}
\label{Kdef}
\eea
where $\cal L$ denotes Lie derivative.
Notice that the constraint (\ref{unit}) that $u^a$ be a unit vector
is built into our initial value formulation by its identification with
the unit normal.

The use of $K_{ab}$ as a variable allows us to write Einstein's equation
in a form that is first order in time.  Similarly we introduce the
quantity $P={{\cal L}_u}\psi $ which allows us to do the same for the
wave equation.

The time evolution vector field takes the form
${t^a}=\alpha {u^a}+{\beta ^a}$.
We use as a radial coordinate $r$ the
length in the radial direction (rather than an area coordinate).  This makes
the spatial metric
\beq
{h_{ab}} = {\partial _a} r {\partial _b} r + {\Phi ^2} {\Sigma _{ab}}
\eeq
where $\Sigma _{ab}$ is the unit two-sphere metric.

The quantities that are evolved are ($\psi ,P,K,{a_r},\Phi$).
All other quantities
are found from these.
In the appendix, equations of motion for these quantities are
derived from the Euler-Lagrange equations (\ref{EFE}-\ref{unit}).
These equations are
\bea
{\partial _t}\psi &=& \alpha P + {\beta ^r}{\partial _r} \psi
\label{dtpsi2}
\\
\nonumber
{\partial _t} P &=& {\beta ^r}{\partial _r} P
\\
&+& \alpha \left [
P K + {a^r}{\partial _r}\psi + {\partial _r}{\partial _r} \psi +
{{2 {\partial _r}\Phi}\over \Phi}{\partial _r} \psi \right ]
\label{dtP2}
\\
\nonumber
{\partial _t} K &=& {\beta ^r} {\partial _r} K +
{\alpha \over 3} {K^2}
\\
\nonumber
&+& {\alpha \over {2 + {c_{13}} +3{c_2}}}
\biggl [ ({c_{14}}-2)({\partial _r}{a_r}+2{a_r}{\partial _r}\Phi/\Phi +
{a_r}{a_r})
\\
&+&2 {P^2}+3(1-{c_{13}}){Q^2} \biggr ]
\label{dtK2}
\\
\nonumber
{\partial _t} {a_r} &=& {\beta ^r} {\partial _r} {a_r} + \alpha \biggl [
\bigl ( {{2K} \over 3} -  Q\bigr ) {a_r}
\\
&+&
{{c_{13}} \over {{c_{14}}(1-{c_{13}})}} P {\partial _r} \psi
- {{c_{123}}\over {{c_{14}}(1-{c_{13}})}} {\partial _r} K \biggr ]
\label{dta2}
\\
{\partial _t} \Phi &=& {\beta ^r} {\partial _r} \Phi + \alpha \Phi
(Q/2-K/3). \label{dtPhi2} \eea
Here $Q = {{K^r}_r}-(K/3)$ is the
trace-free part of the extrinsic curvature. As shown in the
appendix, the quantities $Q,\, \alpha$ and $\beta$ are determined by
\bea \nonumber {\partial _r} Q &=& - {{3 Q} \over \Phi} {\partial
_r} \Phi + {{(1-{c_{13}})}^{-1}} \bigl [
{\textstyle {1\over 3}} (2+{c_{13}}+3{c_2}) {\partial _r} K\\
&-& P {\partial _r} \psi \bigr ]
\label{drQ2}
\\
{\partial _r} \ln \alpha &=& {a_r}
\label{dralpha2}
\\
{\partial _r}{\beta ^r} &=& \alpha (Q + K/3). \label{drbeta2} \eea
The Hamiltonian initial value constraint  is \bea \nonumber {\cal C}
&=& {\partial _r}{\partial _r} \Phi + {{{{({\partial _r} \Phi
)}^2}-1} \over {2 \Phi}} + {c_{14}} {a_r} {\partial _r} \Phi
\\
\nonumber
&+& {\Phi \over 4} \biggl [ {c_{14}} (2 {\partial _r} {a_r} + {a_r}{a_r})
+{P^2} +{{({\partial _r}\psi )}^2}
\\
&+&{\textstyle {3 \over 2}}(1-{c_{13}}){Q^2} -
{\textstyle {1 \over 3}}(2 + {c_{13}}+3{c_2}) {K^2} \biggr ]\nonumber\\
&=&0.
\label{hamreltext}
\eea

For initial data, we choose a moment of time symmetry, so that $P,Q$
and $K$ are zero.  Both $a_r$ and $\psi$ can be freely specified. We
choose $a_r$ to vanish and $\psi$ to be \beq \psi={a_0} \exp \bigl [
- {{({r^2}-{r_0 ^2})}^2}/{s^4} \bigr ] \eeq where ${a_0}, \, {r_0}$
and $s$ are constants.  Thus the scalar field is initially at rest
and is essentially a spherical shell of radius $r_0$, thickness $s$
and amplitude $a_0$.
Then equation (\ref{hamreltext}) is
integrated outwards to find $\Phi$, subject to the
boundary conditions $\Phi=0$ and $\partial_r\Phi=1$ at $r=0$.
The other boundary conditions that need to be imposed
are $\beta^r=0$ and $Q=0$
for smoothness at $r=0$, and $\alpha\rightarrow1$
for a gauge fixing
as $r\rightarrow\infty$.

The evolution proceeds as follows: given
($\psi ,P,K,{a_r},\Phi$) at one time step, equations
(\ref{dralpha2}) and (\ref{drQ2}) are integrated to find $\alpha$
and $Q$.  Then equation (\ref{drbeta2}) is integrated to find $\beta
^r$. Then equations (\ref{dtpsi2}), (\ref{dtP2}), (\ref{dtK2}),
(\ref{dta2}) and (\ref{dtPhi2}) are used to find respectively ($\psi
,P,K,{a_r},\Phi$) at the next time step.  Since $\Phi$ is evolved
using equation (\ref{dtPhi2}) but must also satisfy equation
(\ref{hamreltext}) this means that equation (\ref{hamreltext}) can
be used as a code check in a convergence test.

\section{Numerical results}
\label{Numresults}

The specific
parameters we used for the initial scalar field are ${a_0}=0.15, \,
{r_0}=10$ and $s = 4$.
Initial data with these parameters leads to
gravitational collapse and the formation of a black hole.

In the numerical simulations spatial derivatives were implemented
with standard centered finite differences.  The time evolution
was done with the Iterated Crank-Nicholson (ICN) method.  Kreiss-Oliger
type dissipation was used for stability.  The first grid point was at
$r=0$.  Of the quantities that are evolved by the code, smoothness requires
that ${a_r}$ and $\Phi$ are odd functions
of $r$ and that the other quantities are even.  The code evolves the
quanties on all grid points except the first and last.  The odd quantities
are set to zero on the first grid point, while the even quantities are
set as follows:
\beq
{f_1} = (4{f_2}-{f_3})/3
\eeq
which implements the condition that the derivative of $f$ vanish
at the origin.
Here $f$ denotes any even quantity, and the subscript denotes the
corresponding gridpoint.  At the last gridpoint, $N$, all the evolved
quantities are set to zero with the exception of $\Phi$ which is set
by
\beq
{\Phi _N}=2{\Phi _{N-1}} - {\Phi _{N-2}}
\eeq
which is the appropriate boundary condition for a quantity that
asymptotically satisfies $\partial \Phi /\partial r =1$.

The simulations were run on a SunBlade 2000 in double precision.  The
number $N$ of gridpoints used was 12800 and the initial value of $r$ at the
outermost gridpoint was 80.
\begin{figure}
\includegraphics[scale=0.6]{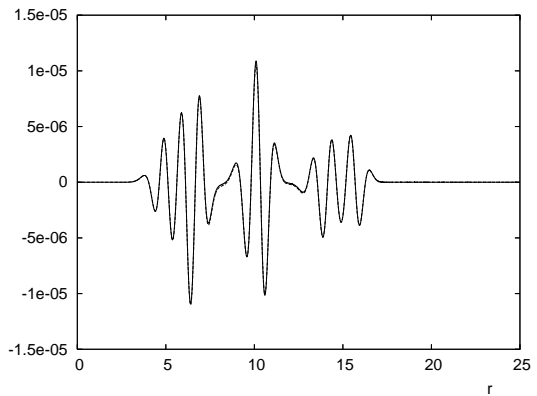}
\caption{\label{Ctest}Plot of $\cal C$ {\it vs} $r$ for a simulation
with 12800 points (solid line) and ${\cal C}/4$ for a simulation
with 6400 points (dashed line).  The agreement of the two curves
demonstrates second order convergence of the code.
The curves are for a time when the
scalar field is in the process of collapsing.}
\end{figure}
Figure (\ref{Ctest}) shows the result of a convergence test of the code.
It is a plot of the constraint quantity $\cal C$ of equation (\ref{hamreltext})
done at a time when the scalar field is in the process of collapsing.  The
solid curve represents $\cal C$ for a simulation with 12800 points,
while the dashed curve represents ${\cal C}/4$ for a simulation with
6400 points.  The fact that the two curves argee demonstrates that the
code is second order convergent.

The simulations show that for the initial data used the collapse
does form a spin-0 horizon (which is contained within an ordinary
horizon, that is a marginally outer trapped surface of the metric
$g_{ab}$).  Furthermore, outside of the spin-0 horizon the metric
and aether field do settle down into a time independent state. In
addition, all the scalar field either falls into the black hole or
escapes, so that near the horizon the time independent state is that
of a pure Einstein-aether black hole with no matter other than the
aether field $u^a$. We are able to follow the time evolution to late
times with our horizon crossing time slices, orthogonal to the
timelike aether field $u^a$, because the lapse $\alpha$ is driven to
zero as the singularity is approached. The reason for this can be
traced to Eqn. (\ref{dralpha2}). Evidently at the singularity the
aether acceleration component $a_r$  goes to positive infinity.

In reference~\cite{Eling:2006ec} the static black holes of
Einstein-aether theory with a regular spin-0 horizon were found
numerically, and it was conjectured that these would be the
endstates of gravitational collapse.  Our simulations verify that
this is indeed what happens.
\begin{figure}
\includegraphics[scale=0.6]{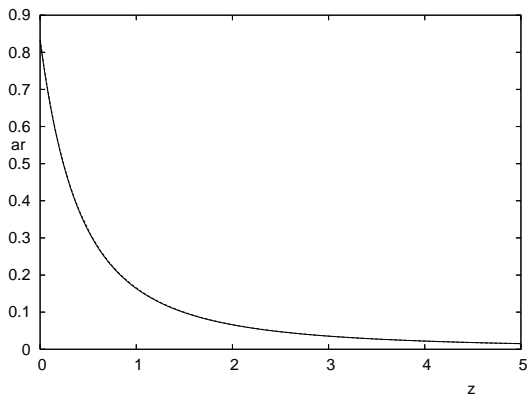}
\caption{\label{arfig}Plot of $a_r$ {\it vs} $z$ for the end state
of the collapse simulation
(solid line) and the static solution found using the method
of \cite{Eling:2006ec} (dashed line)}
\end{figure}
\begin{figure}
\includegraphics[scale=0.6]{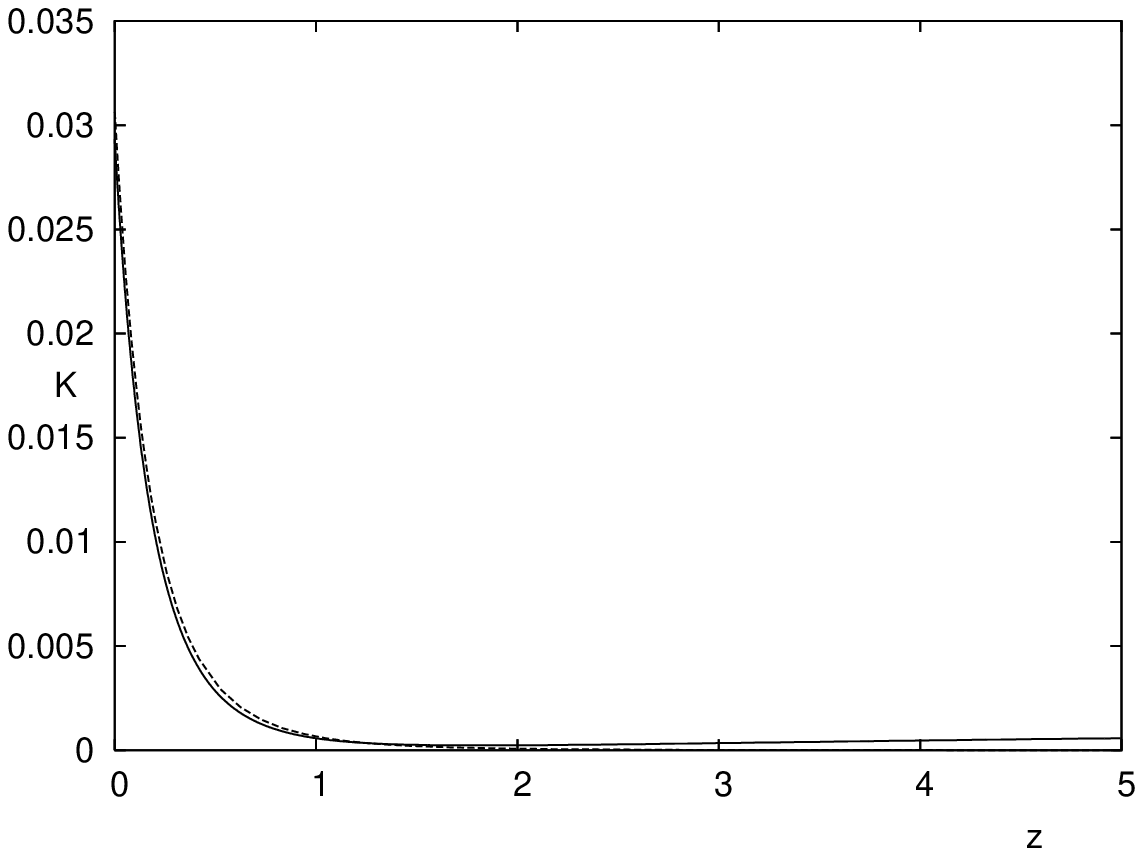}
\caption{\label{Kfig}Plot of $K$ {\it vs} $z$ for the end state
of the collapse simulation
(solid line) and the static solution found using the method
of \cite{Eling:2006ec} (dashed line)}
\end{figure}
\begin{figure}
\includegraphics[scale=0.6]{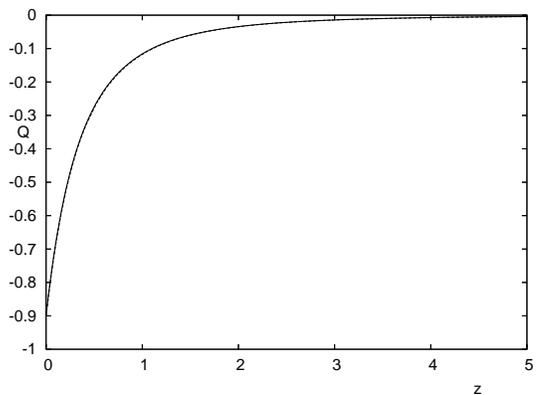}
\caption{\label{Qfig}Plot of $Q$ {\it vs} $z$ for the end state of
the collapse simulation
(solid line) and the static solution found using the method
of \cite{Eling:2006ec} (dashed line)}
\end{figure}
In figures (\ref{arfig}), (\ref{Kfig}) and (\ref{Qfig}) are plotted
respectively the quantities ${a_r}, \, K$ and $Q$ for this
simulation (solid line) and for the static solution found using the
method of~\cite{Eling:2006ec} (dashed line). Here the static
solution is found using the transformed choice of the parameters
$c_i$, and its metric and aether variables are then transformed back
as explained in Section \ref{aetheory} to make the comparison with
the result of the collapse simulation. In all cases the quantities
are plotted as functions of $z=\Phi - {\Phi _0}$ where $\Phi _0$ is
the value of $\Phi$ at the spin-0 horizon. Note that the agreement
is extremely good, so the end result of collapse is indeed the
static black hole solution found using the method of
\cite{Eling:2006ec}.

\begin{figure}
\includegraphics[scale=0.6]{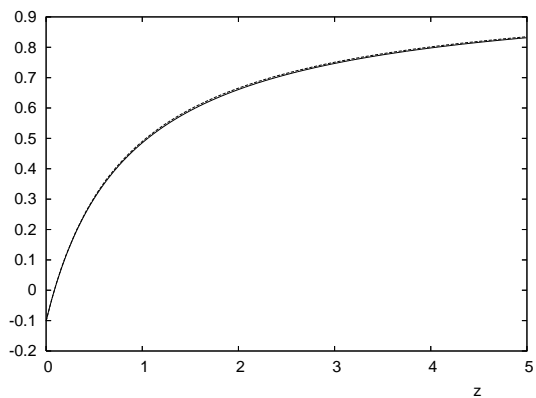}
\caption{\label{grrfig}Plot of ${\nabla ^a}\Phi {\nabla _a} \Phi$
{\it vs} $z$ for the end state of the collapse simulation
(solid line) and the static solution found using the method
of \cite{Eling:2006ec} (dashed line)}
\end{figure}

\begin{figure}
\includegraphics[scale=0.6]{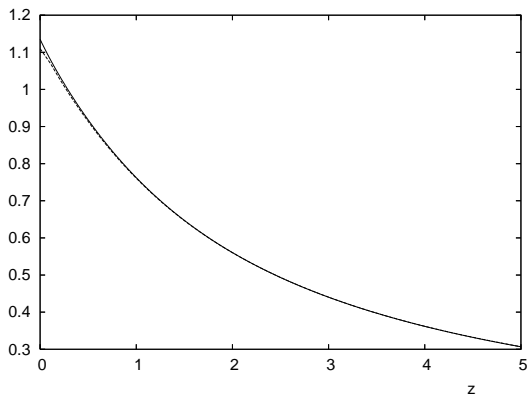}
\caption{\label{boxrfig}Plot of ${\nabla ^a}{\nabla _a}\Phi$
{\it vs} $z$ for the end state of the collapse simulation
(solid line) and the static solution found using the method
of \cite{Eling:2006ec} (dashed line)}
\end{figure}

The quantities
${a_r},\, K$ and $Q$ completely determine the aether field, and since we
are treating spherically symmetric spacetimes where there are no separate
gravitational degrees of freedom, these quantities also completely determine
the metric.  Nonetheless, it is helpful to do a direct comparison of
the metric of the endstate of the collapse simulation and the static metric
found using the method of \cite{Eling:2006ec}.  For a static,
spherically symmetric metric with area radius $\Phi$, the metric is
determined by
the quantities
${\nabla _a}\Phi {\nabla ^a}\Phi$ and ${\nabla _a}{\nabla ^a} \Phi$.  In
figures (\ref{grrfig}) and (\ref{boxrfig}) these quantities are
respectively plotted.
As with figures (\ref{arfig}-\ref{Qfig}) the quantities are plotted
as functions of $z$ with the endstate of the collapse simulation as
a solid line and the static metric found using the method of
reference \cite{Eling:2006ec} as a dashed line.  The agreement
between the curves shows that the metric in the final state of the
collapse simulation agrees with the static metric.

We now present the results of a slightly different simulation that
uses directly a different parameter choice of \cite{Eling:2006ec},
namely ${c_3}={c_4}=0$, and 
$c_2$ chosen so that the spin-0 horizon coincides with the usual
metric horizon. That is, the speed $v_0$ in (\ref{v0}) is set equal
to one, which requires \beq {c_2} = {{-{c_1 ^3}}\over
{2-4{c_1}+3{c_1 ^2}}}. \eeq If $0<c_1<1$ such theories meet all the
linear stability conditions (stable modes, positive energy, and
no vacuum \v{C}erenkov radiation). In
this class the post-Newtonian preferred frame parameter $\a_1$
vanishes identically, but $\a_2$ does not vanish unless $c_1=2/3$ (see
~\cite{Graesser:2005bg,Foster:2005dk} for $\a_{1,2}$). Nevertheless,
since the stability conditions are met, it seems likely that the
behavior of collapse in the theory is qualitatively similar to the
previous case where also $\a_2$ vanishes.
\begin{figure}
\includegraphics[scale=0.6]{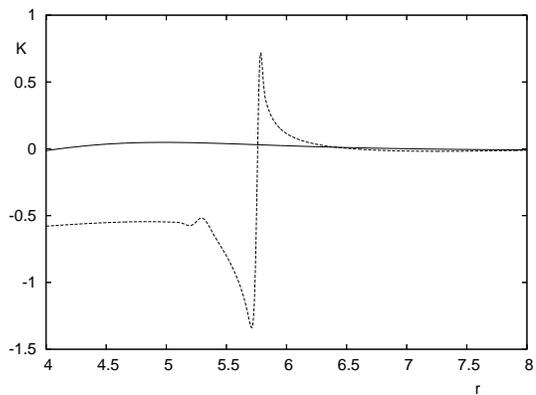}
\caption{\label{c1fig}Plot of $K$ {\it vs} $r$ for simulations
with ${c_1}=0.7$ (solid line) and ${c_1}=0.8$
(dashed line).}
\end{figure}
In~\cite{Eling:2006ec}, static, asymptotically flat black hole
solutions were found in the range ${c_1}\le0.7$, by integrating from
a regular horizon to spatial infinity using a shooting method. The
asymptotically flat boundary condition could not be met for
${c_1}\ge0.8$.

We performed collapse simulations
for the case ${c_1}=0.7$ and the case ${c_1}=0.8$.  In the
${c_1}=0.7$ case a regular horizon forms.
However, in the ${c_1}=0.8$
case, no horizon forms and the evolution seems to become singular,
thus indicating the formation of a naked singularity.
Figure (\ref{c1fig}) Shows the value of $K$ (dashed line) for the
${c_1}=0.8$ simulation at a time shortly before the singularity forms.
For comparison, the same figure shows (solid line) the value of $K$
at the same time for the ${c_1}=0.7$ simulation.
Since the collapse solutions are by construction
asymptotically flat, this
is consistent with what was found for static solutions in
\cite{Eling:2006ec}.

\section{Concluding remarks}

We have examined the process of dynamical collapse of a
spherically symmetric,
self-gravitating scalar field pulse in Einstein-aether theory.
In pure Einstein gravity it is well known that for appropriate initial
conditions such a pulse can form a black hole.
Ours is the first study of the analogous
process in Einstein-aether theory.
In order to carry out the numerical
simulations we developed the initial value formulation
for this theory in spherical symmetry.

Here we have
minimized the initial ``distortion" of the aether field, by setting
its divergence and acceleration to zero. (At the other extreme, one
could study collapse of a pure spherical aether pulse, with no
matter field at all.)  These initial conditions are
not precisely
what would occur in an astrophysical setting, but should be
indicative. We found that for the initial data studied, black holes
with regular metric and spin-0 mode horizons form when the
parameters in the Lagrangian are not too large. For larger
parameters, a singularity occurs at finite area where the spin-0
horizon would have been. These results are consistent with findings
from previous investigation of static solutions~\cite{Eling:2006ec}.

Since regular black holes have almost certainly been ``observed" in
nature, the black hole formation found here indicates that a
strong-field test has been passed by ae-theory, at least for
coupling parameters that are not 
so large that the spin-0 horizon is singular.
Constraints
from the absence of vacuum \v{C}erenkov radiation require that the
spin-0 mode travels at a speed greater than or equal to the ``speed
of light" as seen by matter fields~\cite{Elliott:2005va}, here the
``metric" speed of light. Thus 
a
the singular spin-0 horizon would
necessarily
be
on or inside the matter horizon. Whether it could then have any
observational consequences remains to be determined.

It would also be interesting and astrophysically important to
examine the case of collapse with angular momentum, to a spinning
black hole. The stationary spinning solutions have not yet been
studied however, and to follow the time-dependent collapse would
require the general initial value formulation, without assuming
hypersurface orthogonal aether nor spherical symmetry.  In addition,
axisymmetric codes are much more complicated.

\begin{acknowledgments}
This work was supported in part by the National Science Foundation
under grants  PHY-0601800 through the University of Maryland and
PHY-0456655 through Oakland University.
\end{acknowledgments}

\appendix*
\section{Derivation of the initial value formulation
for spherically symmetric Einstein-aether theory}

In this section we derive the evolution equations for Einstein-aether
theory in spherical symmetry.  To begin with, we use only the hypersurface
orthogonality of $u^a$ and the Euler-Lagrange equations of motion
(\ref{EFE}-\ref{unit}).

Equation (\ref{evolveu}) (the equation of motion for the aether
field) then becomes \beq {c_{14}} \left ( {{\cal L}_u} {a_a} + 2
{K_{ab}}{a^b} - K {a_a} \right ) + {c_{13}}{D^b}{K_{ab}} + {c_2}
{D_a} K = 0. \label{evolvea} \eeq Here ${\cal L}$ denotes the Lie
derivative and $D_a$ is the spatial covariant derivative.

The usual initial value formulation for general relativity requires
that we evolve $h_{ab}$ and $K_{ab}$ as well as the matter fields.
From equation (\ref{Kdef}) we have \beq {{\cal L}_u} {h_{ab}} = - 2
{K_{ab}} \label{evolveh} \eeq which is used to evolve $h_{ab}$.  The
quantity $K$ (the trace of ${K^a}_b$) sasfies the evolution equation
\beq {{\cal L}_u} K = - {D^a}{a_a} - {a^a}{a_a} + {K^{ab}}{K_{ab}} +
{\textstyle {1 \over 2}} {T_{ab}}({h^{ab}} +{u^a}{u^b}).
\label{evolveK} \eeq In addition, the quantity $K_{ab}$ satisfies
the constraint equation \beq {D^b}{K_{ab}} - {D_a} K = -
{{h_a}^b}{u^c}{T_{bc}}. \label{constrainK} \eeq The extrinsic
curvature ${K^a}_b$ can be decomposed into a trace and a trace-free
part.  We will use equation (\ref{evolveK}) to evolve the trace $K$
and will use equation (\ref{constrainK}) to find the trace-free
part.

However, there is a subtlety associated with the Einstein-aether
theory that makes the initial value problem somewhat complicated:
the stress energy tensor contains terms that have second time
derivatives of the aether field and second time derivatives of the
metric.  Thus after writing out the Einstein field equations one
must solve them for these second time derivatives. In particular,
some straightforward but tedious algebra applied to equation
(\ref{aestress}) yields \bea \nonumber {T_{ab}}({h^{ab}}
+{u^a}{u^b}) &=& - ({c_{13}}+3{c_2}){{\cal L}_u} K +
{c_{14}}({D_a}{a^a}+{a_a}{a^a})
\\
&+& 2{P^2} - 2 {c_{13}}{K_{ab}}{K^{ab}} +
{c_{123}} {K^2}
\label{splusrho}
\\
\nonumber
- {{h_a}^b}{u^c}{T_{bc}} &=& - P {D_a} \psi
\\
&-& {c_{14}} \left (  2 {K_{ab}}{a^b} - K {a_a}
+ {{\cal L}_u}{a_a} \right )
\label{tzeroi}
\eea
where $P={{\cal L}_u} \psi $ and ${c_{123}}={c_1}+{c_2}+{c_3}$.
Using equation (\ref{splusrho}) in equation (\ref{evolveK}) and
solving for ${{\cal L}_u} K$ yields
\bea
\nonumber
(2+{c_{13}}+3{c_2}) &{{{\cal L}_u} K}& = ({c_{14}}-2)({D_a}{a^a}+{a_a}{a^a})
+2 {P^2}
\\
&+& 2(1-{c_{13}}){K_{ab}}{K^{ab}} + {c_{123}}{K^2}. \label{evolveK2}
\eea Similarly, equation (\ref{constrainK}), upon substituting the
expression in equation (\ref{tzeroi}), does not directly yield a
constraint since the right hand side of equation (\ref{tzeroi})
contains a term proportional to ${{\cal L}_u}{a_a}$.  However,
eliminating this term using equation (\ref{evolvea}) yields the
following constraint \beq [1-{c_{13}}]{D^b}{K_{ab}}=(1+{c_2}){D_a}K
- P {D_a}\psi \label{constrainK2} \eeq which can be used to rewrite
the equation of motion for $a_a$ as \bea \nonumber {{\cal L}_u}{a_a}
&=& - 2 {K_{ab}}{a^b}+K{a_a} + {{c_{13}}\over
{{c_{14}}(1-{c_{13}})}} P {D_a}\psi
\\
&-& {{c_{123}}\over {{c_{14}}(1-{c_{13}})}} {D_a} K.
\label{evolvea2} \eea

The usual Hamiltonian constraint of general relativity is
\beq
{^{(3)}}R+{K^2}-{K^{ab}}{K_{ab}} = 2 {T_{ab}}{u^a}{u^b}
\label{hamrel}
\eeq
where ${^{(3)}}R$ is the scalar curvature of the spatial metric.  However,
it follows from equation (\ref{aestress}) that
\bea
\nonumber
2 {T_{ab}}{u^a}{u^b} &=& 2 {c_{14}} {D_a}{a^a} + {P^2} + {D_a}\psi {D^a} \psi
\\
&+& {c_{14}} {a_a}{a^a} - {c_2} {K^2} - {c_{13}}{K_{ab}}{K^{ab}}.
\label{aerho} \eea On substituting equation (\ref{aerho}) into
equation (\ref{hamrel}) one finds the constraint \bea \nonumber
{^{(3)}}R &=& {c_{14}} (2 {D_a}{a^a}+{a_a}{a^a}) + {P^2} + {D_a}\psi
{D^a}\psi
\\
&+& (1-{c_{13}}){K_{ab}}{K^{ab}} - (1+{c_2}){K^2}. \label{hamrel2}
\eea

We now impose spherical symmetry.  We use as a radial coordinate $r$
the length in the radial direction (rather than an area coordinate).
This makes the spatial metric \beq {h_{ab}} = {\partial _a} r
{\partial _b} r + {\Phi ^2} {\Sigma _{ab}} \eeq where $\Sigma _{ab}$
is the unit two-sphere metric.  The time evolution vector field
takes the form ${t^a} = \alpha {u^a} + {\beta ^a}$. From the
definition of $P$ we find \beq {\partial _t}\psi = \alpha P + {\beta
^r}{\partial _r} \psi \label{dtpsi} \eeq while from the wave
equation we obtain \beq {\partial _t} P = {\beta ^r}{\partial _r} P
+ \alpha \left [ P K + {a^r}{\partial _r}\psi + {\partial
_r}{\partial _r} \psi + {{2 {\partial _r}\Phi}\over \Phi}{\partial
_r} \psi \right ]. \label{dtP} \eeq

It is helpful to define $Q\equiv {{K^r}_r}-(K/3)$ so that $Q$ is the
component of the trace-free part of ${K^a}_b$. From equation
(\ref{evolveh}), we have the standard result \beq {{\cal L}_t}
{h_{ab}} = - 2 \alpha {K_{ab}} + {{\cal L}_\beta} {h_{ab}}.
\label{evolveh2} \eeq The $rr$ component of equation
(\ref{evolveh2}) yields \beq {\partial _r}{\beta ^r} = \alpha (Q +
K/3) \label{drbeta} \eeq while the $\theta \theta $ component yields
\beq {\partial _t} \Phi = {\beta ^r} {\partial _r} \Phi + \alpha
\Phi (Q/2-K/3). \label{dtPhi} \eeq From the definition of $\alpha$
we have \beq {\partial _r} \ln \alpha = {a_r}. \label{dralpha} \eeq
Equation (\ref{constrainK2}) yields \beq {\partial _r} Q = - {{3 Q}
\over \Phi} {\partial _r} \Phi + {{(1-{c_{13}})}^{-1}} \left [
{\textstyle {1\over 3}} (2+{c_{13}}+3{c_2}) {\partial _r} K - P
{\partial _r} \psi \right ] \label{drQ} \eeq while equations
(\ref{evolvea2}) and (\ref{evolveK2}) yield respectively \bea
\nonumber {\partial _t} {a_r} &=& {\beta ^r} {\partial _r} {a_r} +
\alpha \biggl [ \bigl ( {{2K} \over 3} -  Q\bigr ) {a_r}
\\
&+&
{{c_{13}} \over {{c_{14}}(1-{c_{13}})}} P {\partial _r} \psi
- {{c_{123}}\over {{c_{14}}(1-{c_{13}})}} {\partial _r} K \biggr ]
\label{dta}
\\
\nonumber
{\partial _t} K &=& {\beta ^r} {\partial _r} K +
{\alpha \over 3} {K^2}
\\
\nonumber
&+& {\alpha \over {2 + {c_{13}} +3{c_2}}}
\biggl [ ({c_{14}}-2)({\partial _r}{a_r}+2{a_r}{\partial _r}\Phi/\Phi +
{a_r}{a_r})
\\
&+&2 {P^2}+3(1-{c_{13}}){Q^2} \biggr ]. \label{dtK} \eea The
Hamiltonian constraint (equation (\ref{hamrel2})) becomes the
vanishing of the quantity $\cal C$ where \bea \nonumber {\cal C} &=&
{\partial _r}{\partial _r} \Phi + {{{{({\partial _r} \Phi )}^2}-1}
\over {2 \Phi}} + {c_{14}} {a_r} {\partial _r} \Phi
\\
\nonumber
&+& {\Phi \over 4} \biggl [ {c_{14}} (2 {\partial _r} {a_r} + {a_r}{a_r})
+{P^2} +{{({\partial _r}\psi )}^2}
\\
&+&{\textstyle {3 \over 2}}(1-{c_{13}}){Q^2} - {\textstyle {1 \over
3}}(2 + {c_{13}}+3{c_2}) {K^2} \biggr ]. \label{hamrel3} \eea

\end{document}